\documentclass[structabstract]{aa}
\usepackage{graphicx}
\usepackage{txfonts}
\def\ltsim{\raise 2pt \hbox {$<$} \kern-1.1em \lower 4pt \hbox {$\sim$}}
\def\gtsim{\raise 2pt \hbox {$>$} \kern-1.1em \lower 4pt \hbox {$\sim$}}

\begin{document}
%
%
%

\title{The very steep spectrum radio halo in Abell 697}

\author{G.~Macario\inst{1,}\inst{2}, 
T.~Venturi\inst{1}, 
G.~Brunetti\inst{1},
D.~Dallacasa\inst{1,}\inst{2}, 
S.~Giacintucci\inst{3,}\inst{1},
R.~Cassano\inst{1}\and
S.~Bardelli\inst{4} \and
R.~Athreya\inst{5}}
\institute
{
INAF -- Istituto di Radioastronomia, via Gobetti 101, I-40129, Bologna, Italy
\and
Dipartimento di Astronomia, Universit\'a di Bologna, via Ranzani 1, I--40127,
Bologna, Italy
\and
Harvard--Smithsonian Centre for Astrophysics,
60 Garden Street, Cambridge, MA 02138, USA
\and
INAF--Osservatorio Astronomico di Bologna, via Ranzani 1, I--40127, Bologna,
Italy
\and
Indian Institute of Science Education and Research, Sutarwadi Road, Pashan, 
Pune 411021, INDIA 
}
%
\date{Received 00 - 00 - 0000; accepted 00 - 00 - 0000}
%
\titlerunning{The very steep spectrum radio halo in Abell 697}
\authorrunning{Macario et al.}
\abstract
{}
{In this paper we present a detailed study of the giant radio halo
in the galaxy cluster Abell 697, with the aim to constrain its origin and
connection with the cluster dynamics.}
{We performed high sensitivity GMRT observations at 325 MHz, which showed
that the radio halo is much brighter and larger at this frequency, compared to
previous 610 MHz observations. In order to derive the integrated spectrum 
in the frequency range 325 MHz--1.4 GHz,  we re--analysed 
archival VLA data at 1.4 GHz and made use of proprietary GMRT data at 610 MHz.} 
{Our multifrequency analysis shows that the total radio spectrum of the
giant radio halo in A\,697 is very steep, with 
$\alpha_{\rm~325 MHz}^{\rm~1.4 GHz} \approx 1.7-1.8$. 
Due to energy arguments, a hadronic origin of the halo is disfavoured
by such steep spectrum.
Very steep spectrum halos in merging clusters are predicted in the case
that the emitting electrons are accelerated by turbulence, observations
with the upcoming low frequency arrays will be able to test these
expectations.}
{}
\keywords
{radiation mechanism: non-thermal -- galaxies: clusters: general -- 
galaxies: clusters: individual: A\,697}

\maketitle
%

\section{Introduction}\label{sec:intro}

Radio halos are diffuse low surface brightness radio sources extended on 
Mpc scales, 
observed in the central regions of a fraction of X--ray luminous (i.e. 
massive) galaxy clusters. Their synchrotron emission is not associated with 
individual galaxies, but it origins from the non--thermal component of the 
intracluster medium (ICM). They provide the most important piece of
evidence for the
presence of relativistic ($\sim$GeV) electrons and $\mu$G magnetic fields in 
galaxy clusters (see Ferrari et al. \cite{ferrari08} for a recent review). 
To date radio halos have been found from large sky surveys (such as the 
Northern VLA Sky Survey, NVSS, Condon et al. \cite{condon98} and the 
Westerbork Northern Sky Survey, WENSS, Rengelink et al. \cite{rengelink97}; 
i.e. Giovannini et al. \cite{giovannini99}; Kempner \& Sarazin 
\cite{kempner01}), and from 
pointed observations of individual clusters, mostly carried out at 
1.4 GHz (Govoni et al. \cite{govoni01}; Bacchi et al. \cite{bacchi03}; and
the recent compilation by Giovannini et al. \cite{giovannini09}) and at
610 MHz (the GMRT Radio Halo survey, Venturi et al. \cite{venturi07} and 
\cite{venturi08}, hereinafter VGB07 and VGD08).
They have steep radio spectra, with $\alpha \sim 1.2-1.4$ 
($S\propto \nu^{-\alpha}$) from integrated flux density measurements.
\newline
\indent
The origin of radio halos is a long--standing problem, since the radiative 
life--time of the relativistic electrons responsible for their synchrotron 
emission is much shorter than the particle diffusion time required to cover 
their Mpc extent (Jaffe \cite{jaffe77}).
This requires some form of particle re--acceleration. 
So far two main models have been proposed to explain the origin of
radio halos: 
(1) the \textquotedblleft{re--acceleration model}\textquotedblright, 
whereby radio halos originate from the in--situ re--acceleration of 
pre--existing electrons by magneto--hydrodynamic (MHD) turbulence injected
in the ICM during cluster mergers (Brunetti et al. \cite{brunetti01}; 
Petrosian \cite{petrosian01}; Brunetti \& Blasi \cite{brunetti05}); 
(2) \textquotedblleft{secondary electron models}\textquotedblright, predicting 
that relativistic electrons are injected in the cluster volume by hadronic 
collisions between relativistic cosmic rays and the thermal protons of the ICM 
(e.g. Dennison \cite{dennis80}; Blasi \& Colafrancesco \cite{blasi99}).\\
Recent radio observations provide indirect evidence for turbulent 
acceleration in the ICM (e.g. Brunetti et al. \cite{a521nature08}).
Observational support to the ``re--acceleration scenario'' is also given 
by the growing 
evidence for the connection between cluster mergers and radio halos, based on 
high sensitivity radio and X--ray 
observations (Buote \cite{buote01}; Govoni et al. \cite{govoni04}; VGD08 
and references therein; Giacintucci et al. \cite{giaci09}; 
Cassano \cite{cassano09}), and by
constraints to the evolution of radio halos (Brunetti et al. \cite{brunetti09}).

The analysis of the radio spectrum of halos is important to address the 
question of their 
origin. The turbulent re--acceleration model provides the unique expectation 
of spectra much steeper than those found to date, as a consequence 
of less energetic merger events (e.g. Cassano \cite{cassano09}). 
The detection of radio halos with very steep spectrum ($\alpha >$ 1.6) would 
be a major piece of evidence in support of this scenario, and at the same
time it would disfavour a secondary origin of the electrons, which requires a 
very large proton energy budget (e.g. Brunetti \cite{brunetti04}). 
The prototype of these sources, which we refer to as \textit{ultra steep 
spectrum radio halos} (hereinafter USSRH), 
has been recently discovered in the merging cluster A\,521 and has a spectrum 
with $\alpha\sim$1.9 
(Brunetti et al. \cite{a521nature08}; Dallacasa et al. \cite{dallacasa09}).
\\
In this paper we present a detailed study of the merging cluster A\,697, 
which hosts a candidate steep spectrum radio halo. 
Our work is based on proprietary GMRT radio 
observations at 610 and 325 MHz, and data from the VLA public archive at 
1.4 GHz. 

A\,697 is a rich and massive galaxy cluster located at z=0.282, belonging
to the Abell, Corwin and Olowin catalogue (ACO catalogue, 
Abell et al. \cite{abell89}).
The cluster is hot and luminous in the X--ray band, and is part of the ROSAT 
Brightest Cluster Sample (BCS; Ebeling et al. \cite{ebeling98}). 
Its general properties are summarized in Table \ref{tab:cluster_prop}
\footnote{We adopt the $\Lambda$CDM cosmology with 
H$_0$=70 km s$^{-1}$ Mpc$^{-1}$, $\Omega_m=0.3$ and $\Omega_{\Lambda}=0.7$.  
At the redshift of A\,697 (z=0.282), this cosmology leads to a linear 
scale of $1^{\prime \prime}=4.26$ kpc.}, which provides: coordinates;
morphological classification of the cluster and richness; redshift z;
cluster velocity dispersion $\sigma_v$ (from optical spectroscopy);
X--ray luminosity L$_{\rm X}$ (taken from the BCS catalogue); 
virial mass M$_V$ and the corresponding virial radius R$_V$. 
The presence of diffuse cluster--scale radio emission in A\,697 was first 
suggested in Kempner \& Sarazin (\cite{kempner01}) by inspection of 
the NVSS and the WENSS, 
and further confirmed by observations at 610 MHz with the Giant Metrewave 
Radio Telescope (GMRT) as part of the GMRT Radio Halo Survey (VGB07 and VGD08). 
From those observations the extended radio emission at the cluster centre has 
been unambiguously classified as a 
giant\footnote{Linear size \gtsim~ 1 Mpc h$^{-1}_{\rm 50}$.}
radio halo. 
%
%
\begin{table}
\label{tab:cluster_prop}
\caption[]{General properties of the galaxy cluster A\,697.}
\begin{center}
\footnotesize
\begin{tabular}{ll}
\hline\noalign{\smallskip}
RA$_{J2000}$  &  08h 42m 53.3s \\
DEC$_{J2000}$ & $+$36$^{\circ}$ 20$^{\prime}$ 12$^{\prime \prime}$ \\
Bautz--Morgan Class & II--III \\
Richness & 1 \\
z & 0.282  \\
$\sigma_v$  & 1334 km s$^{-1}$ (a) \\
L$_{\rm X[0.1-2.4 keV]}$  & 10.57 $\times$ 10$^{44}$ erg s$^{-1}$\\
M$_{\rm V}$ &  2.25 $\times$ 10$^{15}$ M$_{\odot}$ (b)\\
R$_{\rm V}$ &  2.90 Mpc (b)\\
\hline\smallskip
\end{tabular}
\end{center}
Notes to Table \ref{tab:cluster_prop}: (a): Girardi et al. \cite{girardi06}; 
(b): Estimated from the L$_X$--M$_V$ relation, see Eq. 6 and 7 in Cassano 
et al. (\cite{cassano06}).
\end{table}
%
%

The paper is organized as follows. In Sect. \ref{sec:GMRT325} we present
the new 325 MHz GMRT data; in Sect. \ref{sec:spectrum} and 
\ref{sec:injections} we derive the integrated spectrum and discuss
the sources of uncertainty in the flux density measurements; 
in Sect. \ref{sec:discussion} we discuss the
origin of the A\,697 radio halo in the framework of the turbulent 
re--acceleration model. Summary and conclusions are given in \ref{sec:results}.

\section{GMRT radio data at 325 MHz}\label{sec:GMRT325}

\subsection{Radio observations and data reduction}\label{sec:obs}

The main characteristics of the 325 MHz observations are summarised in Table 
\ref{tab:obs}, which reports: 
observing date, frequency, total bandwidth, total time on source, 
synthesized half power 
beamwidth (HPBW), rms level (1$\sigma$) at full resolution and u--v range 
of the full dataset. 

The observations were carried out using simultaneously the upper and lower side
bands (USB and LSB, respectively), for a total observing bandwidth of 32 MHz. 
The default spectral--line observing mode was performed, with 128 channels for 
each band and a spectral resolution of 125 kHz/channel. 
The USB and LSB datasets were calibrated and analyzed separately using the NRAO
Astronomical Image Processing System package(AIPS). 
3C\,147 and 3C\,286 were used as primary calibrators, and were observed 
respectively at the beginning and at the end of the observing run. The point 
source 0735+331 was used as phase calibrator. 
\\
Due to the considerably lower quality of the LSB (strong RFI residuals),
only the USB data were used to produce the final images presented and
analysed in this paper.
\\
The very large field of view of the GMRT at 325 MHz (primary beam
$\sim 1.8^{\circ}$) required the implementation of the wide--field imaging 
technique in each step of the self--calibration process, in order to 
account for the non--planar nature of the sky. 
We covered a field of view as large as $\sim 2.7\times 2.7$ square degrees
with 25 facets, to include possible strong sources located beyond the primary 
lobe, conventionally cut at 5\%. 
We initially self--calibrated the longest baselines using only the point
sources in the field, then we progressively included short baselines and
resolved radio sources. Finally we included also the emission from the
radio halo. Note that only phase self--calibration was applied.
\\ 
Even though only half of the full dataset was usable, the quality of the final 
image is very good: the rms ranges from $\sim 45$ $\mu$Jy b$^{-1}$ 
to $\sim 55$ $\mu$Jy b$^{-1}$. Despite the relatively low frequency, 
confusion in this image is negligible, due to the arcsecond resolution.
\\
The residual amplitude errors on each individual antenna are $\lesssim5\%$. 
On the basis of this result we can conservatively assume that the absolute
flux density calibration is within 5\%. Such value accounts for the
uncertainty in the calibrator flux density scale as well.

%
\begin{table*}[htbp!]
\caption[]{Summary of the GMRT radio observations.}
\begin{center}
\footnotesize
\begin{tabular}{cclcccc}
\hline\noalign{\smallskip}
 Observation & $\nu$ & $\Delta \nu$  & t  & HPBW, p.a. & rms & {\it u--v} range\\ 
   date & (MHz)& (MHz) & (min) & (full array , $^{\prime \prime}\times^{\prime \prime}$, $^{\circ}$)&  ($\mu$Jy b$^{-1}$) & (k$\lambda$)\\ 
\noalign{\smallskip}
\hline\noalign{\smallskip}
  17 Jan 2007 & 325 & 32(16)$^{*}$ &  480 & 10.0$\times$9.1, -64 & 45 & $\sim$0.08-25 \\
\noalign{\smallskip}
\hline\noalign{\smallskip}
\end{tabular}
\end{center}
Notes to Table \ref{tab:obs}: $^{*}$ The observations were performed using a 
total bandwidth of 32 MHz (USB+LSB), but only the USB dataset was used for the 
analysis (see Sect. \ref{sec:obs}) 
\label{tab:obs}
\end{table*}
%
%
%

\subsection{The field}\label{sec:a697_field}

\begin{figure*}[htbp]
\centering
\includegraphics[angle=0,scale=0.75]{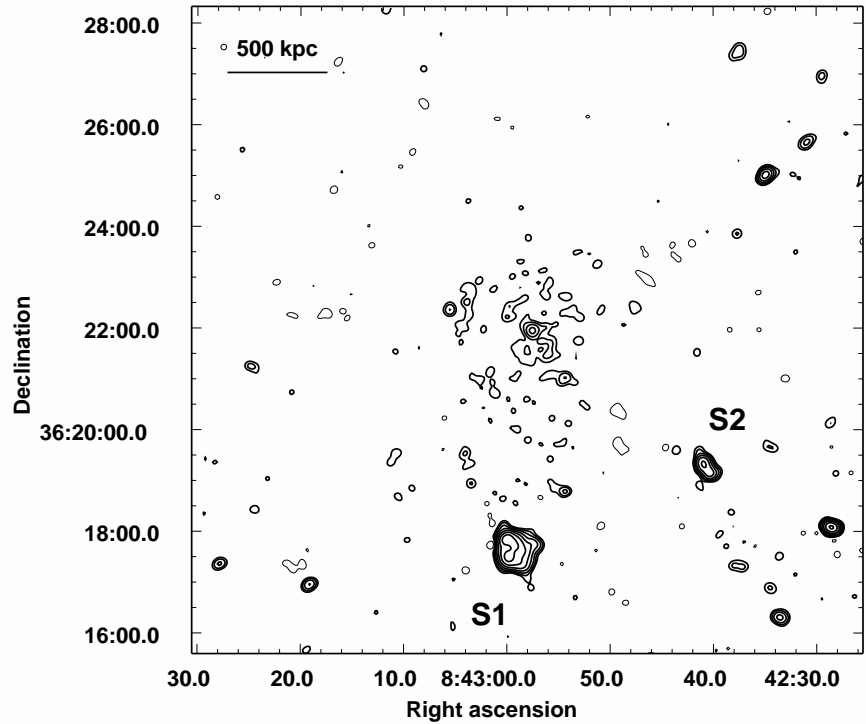}
\caption{GMRT 325 MHz radio contours of the inner 
$12^{\prime}\times 12^{\prime}$ region centered on A\,697.
The 1$\sigma$ level in the image is 45 $\mu$Jy b$^{-1}$.
Contours are spaced by a factor 2 starting from 
5$\sigma=\pm$ 0.225 mJy b$^{-1}$. 
The HPWB is $10.0^{\prime\prime} \times 9.1^{\prime\prime}$, p.a. $-64^{\circ}$. 
The labels S1 and S2 indicate the two extended radio galaxies in the Southern 
part of the cluster.}
\label{fig:a697_field}
\end{figure*}

In Fig. \ref{fig:a697_field} we report the 325 MHz GMRT full resolution 
contours of the central $12^{\prime}\times 12^{\prime}$ ($\sim 3\times 3$
Mpc$^{2}$) portion of the A\,697 field. The region corresponds to half the 
cluster virial radius ($R_V = 2.9$ Mpc, Table \ref{tab:cluster_prop}). 
The most prominent features at this resolution are the two extended radio 
galaxies, labelled as S1 and S2 in Fig. \ref{fig:a697_field}, located South of 
the cluster centre. 
In addition, the diffuse radio emission 
associated with the radio halo is clearly visible around the cluster centre. 

%

\subsection{The radio halo}\label{sec:a697_halo}
%
\begin{figure*}
\centering
\includegraphics[angle=0, scale=1.2]{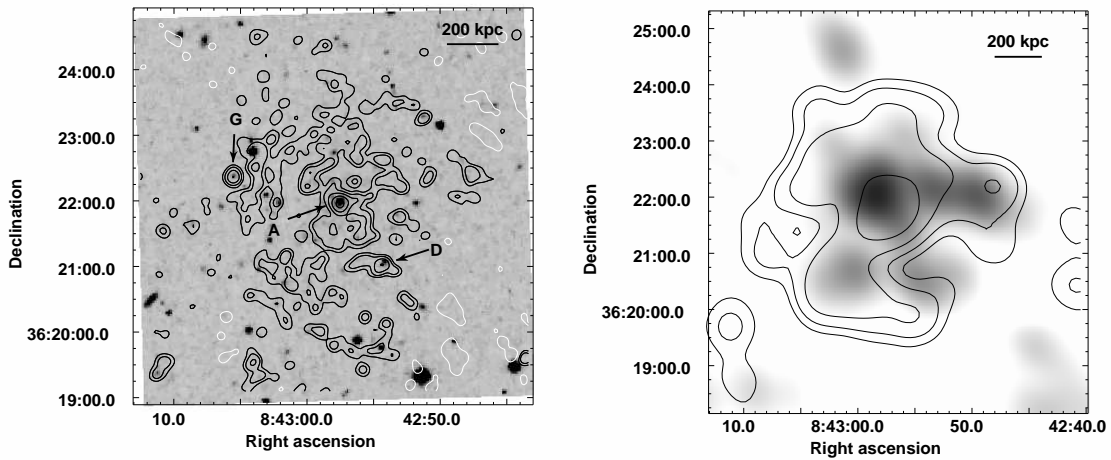}
\caption{{\it Left} -- Full resolution GMRT 325 MHz contours of the
central region of A\,697, superposed to the Second Palomar Sky Survey 
(POSS--2) optical image. The image is the same as Fig. 1  
($10.0^{\prime\prime} \times 9.1^{\prime\prime}$, p.a. $-64^{\circ}$, 
1$\sigma$ level is 0.045 mJy b$^{-1}$) and contours are spaced by a 
factor of 2, starting from  $\pm$ 0.135 mJy b$^{-1}$ (3$\sigma$). 
Individual radio sources are labelled by letters A, D and G (as in VGD08, 
Fig. 2). {\it Right} -- Low resolution GMRT image at 325 MHz of the radio 
halo (obtained after subtraction of the individual sources, see Sect. 2.3) 
overlaid on the GMRT 610 MHz image (grey scale). The resolution of the 
325 MHz image is 
$46.8^{\prime\prime} \times 41.4^{\prime\prime}$, p.a. $80^{\circ}$, and the 
1$\sigma$ noise level is 0.15 mJy b$^{-1}$.
Logarithmic contours are shown, starting from $\pm$0.45 mJy b$^{-1}$.
The 610 MHz image has a resolution of 
$46.4^{\prime\prime} \times 35.9^{\prime\prime}$, p.a. $42^{\circ}$, and the 
1$\sigma$ noise level is 0.05 mJy b$^{-1}$}.
\label{fig:a697_halo}
\end{figure*}
%
\begin{table*} 
\caption[]{Individual radio galaxies at the center of A\,697.}
\begin{center}
\begin{tabular}{lcccc}
\hline\noalign{\smallskip}
Radio     & Radio position  & S$_{325\, \rm MHz}$        &  Optical Position  &  z  \\ 
 source   & RA$_{J2000}$ \& DEC$_{J2000}$ &  mJy  &  RA$_{J2000}$ \& DEC$_{J2000}$  &   \\   
\noalign{\smallskip}
\hline\noalign{\smallskip}
A & 08 42 57.70  +36 22 01 & 4.63$\pm$0.23 & 08 42 57.55 +36 22 00 &  0.281 \\
D & 08 42 54.54  +36 21 05 & 1.97$\pm$0.10 & 08 42 54.36 +36 21 03 &  0.274 \\
G & 08 43 05.65  +36 22 25 & 1.36$\pm$0.07 & 08 43 05.50 +36 22 24 &  0.267 \\
\hline
\end{tabular}
\end{center}
\label{tab:sources}
\end{table*}

\begin{figure*}[htbp]
\centering
\includegraphics[angle=0, scale=0.9, bb=200 330 400 550]{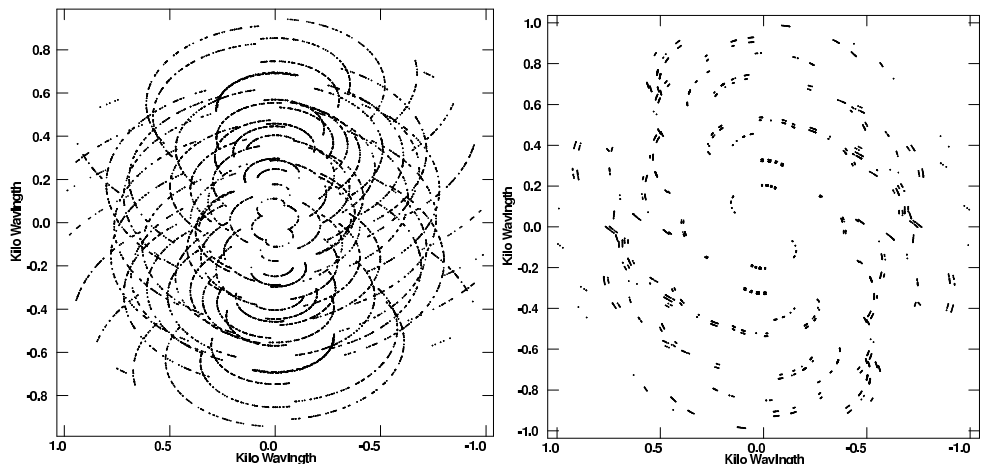}
\vspace{0.5cm}
\caption{Comparison between the inner portion of the u--v plane sampled by the 
GMRT observations at 325 MHz (\textit{left panel}) and  at 610 MHz 
(\textit{right}) from which the flux density measurements of the halo has been 
taken. Only baselines shorter than 1 k$\lambda$ have been plotted.}
\label{fig:uvcoverage}
\end{figure*}

Fig. \ref{fig:a697_halo} zooms into the central cluster region.
The left panel shows the full resolution 325 MHz contours (starting from 
$\pm3\sigma$), overlaid on the red optical frame from the Digitized Palomar 
Sky Survey (DSS--2). Beyond the individual radio sources with optical 
counterpart, diffuse emission is clearly visible at the cluster centre.
Three discrete radio sources embedded in the halo 
emission are optically identified with cluster members (Girardi et al. 
\cite{girardi06}, hereinafter G06). We labelled  them A, D and G, following the
same notation used for the corresponding sources identified in the GMRT 610 MHz
full resolution image (left panel of Fig. 2, see also VGD08). In Table 
\ref{tab:sources} we report their radio position and flux density at 325 
MHz (obtained from Gaussian fits in the full resolution 
image), together with their optical identification and redshift (from G06). 
Source A is identified with the central cD galaxy. 
\\
In order to properly image the diffuse emission of the radio halo, the 
three sources A, D and G, whose total flux density amounts to 
S$_{\rm 325~MHz}= 7.96$ mJy,  were subtracted from the u--v data. 
The final ``subtracted'' dataset was used to produce images of the radio halo 
at various angular resolutions, tapering the u--v data by means of the 
parameters {\tt robust} and {\tt uvtaper} in the task IMAGR.
\\
In the right panel of Fig. \ref{fig:a697_halo} we report the 325 MHz 
contours of the radio halo at the resolution of 
$ 46.8^{\prime\prime} \times 41.4^{\prime\prime}$, overlaid on a GMRT 610 MHz 
image of similar resolution and obtained with a comparable weighting
scheme ($46.4^{\prime\prime} \times 35.9^{\prime\prime}$; grey scale, see also 
VGD08). The first contour corresponds to the $\pm3\sigma$ significance level.
\newline
\noindent
The radio halo of A\,697 is very extended at 325 MHz, with a largest angular 
size (LAS) of $\sim 5.1^{\prime}$, corresponding to a largest linear size (LLS) 
of $\sim$ 1.3 Mpc h$_{\rm 70}^{-1}$. Its overall morphology is regular 
and symmetric, and more extended than imaged at 610 MHz. 
The central $\sim 1^{\prime}$ is similar at both
frequencies, and the bright feature well visible at 610 MHz in the western 
part of the halo is almost coincident with a similar structure at 325 MHz. 
\\
The total integration time of the 610 MHz observations is 
much shorter than that at 325 MHz (see section \ref{sec:obs}), and the 
u--v coverage of the short spacings worse. 
In Fig. \ref{fig:uvcoverage} we report a comparison between the inner portions 
of the u--v plane at 325 and 610 MHz (left and right panel respectively); 
visibilities relative to baselines shorter than 1 k$\lambda$ are 
shown\footnote{The angular size of the halo ($\sim 5^{\prime}$) is sampled by 
visibilities corresponding to baselines shorter than $\sim 0.7$ k$\lambda$.}.  
For this reason, a detailed spectral index imaging was not carried
out.
\\
The total flux density of the radio halo at 325 MHz is S$_{\rm 325 MHz} = 47.3 $
mJy (see Sect. \ref{sec:injections} for a detailed discussion on the 
uncertainties associated with this value); it was obtained by integrating 
the low resolution image (Fig. \ref{fig:a697_halo}, right panel) over 
the region covered by the $\sim3\sigma$ contour. The corresponding total radio 
power is logP$_{\rm 325~MHz}$ (W/Hz)= 25.07. 

%

\section{The integrated radio spectrum of the halo}\label{sec:spectrum}

In order to derive the spectrum of the giant radio halo in A\,697 with
at least three data points, we complemented the GMRT flux density values
at 610 MHz and 325 MHz with archival 1.4 GHz VLA--C public data.

\subsection{VLA archive data at 1.4 GHz}\label{sec:halo_spec_1400}

\begin{figure}[htbp]
\begin{center}
\includegraphics[angle=0,scale=0.48]{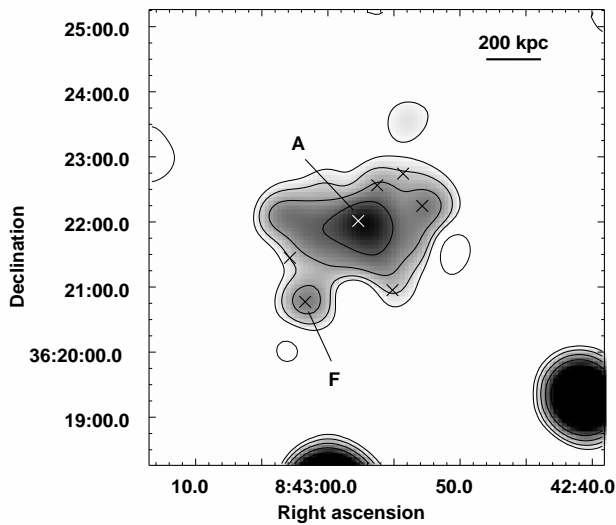}
\caption{Low resolution VLA--C 1.4 GHz image of the radio halo in A\,697.
The 1$\sigma$ level in the image is 25 $\mu$Jy b$^{-1}$.
Contours are spaced by a factor 2 starting from 3$\sigma=\pm$ 
0.075 mJy b$^{-1}$. The restoring beam is $35.0^{\prime\prime} \times 
35.0^{\prime\prime}$, p.a. $0^{\circ}$. Crosses mark the position of the radio 
sources embedded in the halo emission.
}
\label{fig:halo_1400}
\end{center}
\end{figure}

\begin{figure*}[htbp]
\begin{center}
\includegraphics[angle=0,scale=0.98]{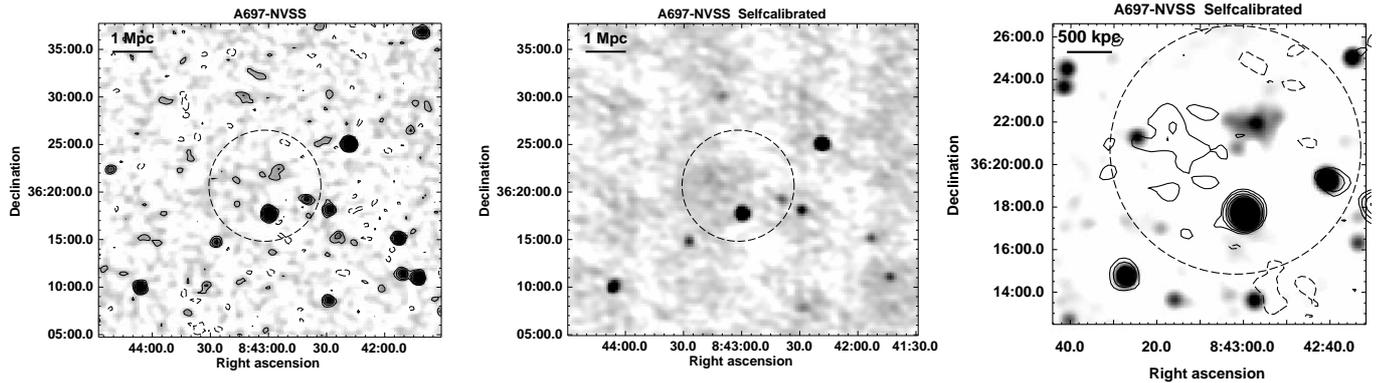}
\caption
{
\textit{Left Panel:} Radio contours of the 
A\,697 field from the public 1.4 GHz NVSS image (grey scale).
The 1$\sigma$ level is 0.45 mJy b$^{-1}$. Contour levels are spaced by a factor
2, starting from $\pm1$ mJy b$^{-1}$; the resolution is 
$45^{\prime\prime}\times45^{\prime\prime}$, p.a 0$^{\circ}$. \textit{Central Panel:} 
1.4 GHz image (grey scale) of the NVSS pointing containing A697 
after new calibration. The 1$\sigma$ level is 0.3 mJy b$^{-1}$. 
The resolution is $45^{\prime\prime}\times45^{\prime\prime}$. 
\textit{Right Panel:} Recalibrated NVSS image (contours) overlaid on the
VLA--C array 1.4 GHz image (same as Fig. 4). The first contour is 
$\pm$0.3 mJy b$^{-1}$, contours are spaced by a factor of 2. The dashed
circle in each panel highlights the portion of the field referred to in
Sect. 3.1.}
\label{fig:NVSS_1400}
\end{center}
\end{figure*}

We re--analyzed archival VLA--C observations at 1.4 GHz 
(Obs. Id. AJ0252). These are short observations centered on 
A\,697, with total integration time of $\sim 50$ minutes and u--v range 
0.3--15 k$\lambda$.
After the usual a--priori calibration, the final images were obtaind after a 
few iterations of phase--only self--calibration.
\\
We successfully detected diffuse emission at the cluster centre.
In order to account for the contribution of the point sources in the halo
region, we produced an image using only baselines longer than 
3 k$\lambda$, to rule out the contribution of the halo emission
(resolution $\sim 9^{\prime\prime}\times11^{\prime\prime}$).
The $1\sigma$ rms is $\sim12~\mu$Jy b$^{-1}$. 
In the central halo region we identified 7 point sources, with peak flux 
density exceeding 5 times the rms level; two of them were already found at 610 
MHz, and labelled A (the central cD) and F in VGD08. We fitted all sources 
with individual Gaussians, and obtained a total contribution to the 
integrated flux density of S$_{\rm 1.4~GHz}=0.97$ mJy. 
In Fig. \ref{fig:halo_1400} we show a low resolution image of the radio 
halo at 1.4 GHz (restoring beam of $35^{\prime\prime}\times35^{\prime\prime}$)
for comparison with the 325 MHz and 610 MHz images. The crosses mark the 
position of the embedded radio point sources. Note that the field shown
is the same as the right panel of Fig. 2.
\\
At 1.4 GHz the radio halo appears considerably smaller than in the GMRT images. 
It is elongated in the East--West direction, in agreement with the brightest 
part of the 610 MHz emission. 
The LAS is $\simeq 190^{\prime\prime}$, corresponding 
to a LLS $\simeq 810$ kpc h$_{\rm 70}^{-1}$. 
Its total flux density is $S_{\rm 1.4~GHz} = 3.7$ mJy, 
corresponding to a radio power logP$_{\rm 1.4~GHz}$ (W/Hz)=23.95.
This value was obtained by integrating this image over the same area covered 
by the radio halo at 325 MHz, and after subtraction of the total contribution 
of the embedded point sources.\\
Hints of diffuse emission at the centre of A\,697 are visible on the NVSS 
1.4 GHz image (Fig. \ref{fig:NVSS_1400}, left panel). However, after an 
accurate inspection we concluded that this image is affected by
fringe residuals (see Giacintucci \cite{giaciPHD07}). 
For this reason, the NVSS pointing containing the A\,697 field was imaged 
after a new calibration. We found out that one IF was 
affected by strong interference, and it was necessary to remove it from the 
self--calibration and imaging process. The new image is reported in the 
central panel of Fig. \ref{fig:NVSS_1400}: indeed, no extended emission is 
detected at the cluster centre at the noise level of 
$\sim$0.3 mJy b$^{-1}$. 
Our re--analysis confirmed that the extended structure visible in the public 
NVSS image is coincident with a peak of a residual fringe, which crosses 
the image along the NW--SE direction, i.e. the same direction of 
the structure itself. Some residual patterns are still visibile in the
recalibrated image at the 1$\sigma$ level. The right panel of 
Fig. \ref{fig:NVSS_1400} reports the recalibrated NVSS image (contours)
overlaid on the 1.4 GHz VLA--C image (grey scale). The non--detection
on the recalibrated NVSS image is consistent with the VLA--C image,
given the rms, peak and restoring beams in the two images (see figure
caption). We integrated the flux density on the recalibrated NVSS image
over the same sky portion covered by the radio halo at 1.4 GHz, and 
measured 2.4 mJy. Considering that the contribution of point sources in
this region is 0.97 mJy (see above), we can conclude that the flux density
of the radio halo on the NVSS amounts to $\simeq$ 1.4 mJy.

\subsection{Integrated radio spectrum}\label{sec:halo_tot_spec}

The observed integrated spectrum of radio halos may be affected by two main
factors. In particular, {\it (1)} the contribution to the flux density of 
embedded individual radio sources, and {\it (2)} the u--v coverage at short 
spacings.
\\
Embedded individual sources are most likely of AGN origin, whose spectrum
is flatter than radio halos. An inaccurate or inefficient
subtraction of these sources may introduce a scatter of the flux density 
measurements at the various frequencies. On the other hand, the u--v 
coverage at short spacings may affect the detection of large scale 
structure. 
\\
We dealt with point {\it (1)} in Sect. 2.3 and 3.1. Point {\it (2)} will be
the focus of Sect. 4.

All the available flux density measurements of the radio halo are reported in 
Table \ref{tab:fluxes}, along with the angular resolution of the images used 
for the measurements. 
The source is undetected on the Very Low--Frequency Sky Survey (VLSS), most 
likely due to the very poor sensitivity of the A\,697 field.

\begin{figure}[htbp]
\centering
\includegraphics[angle=0,scale=0.4]{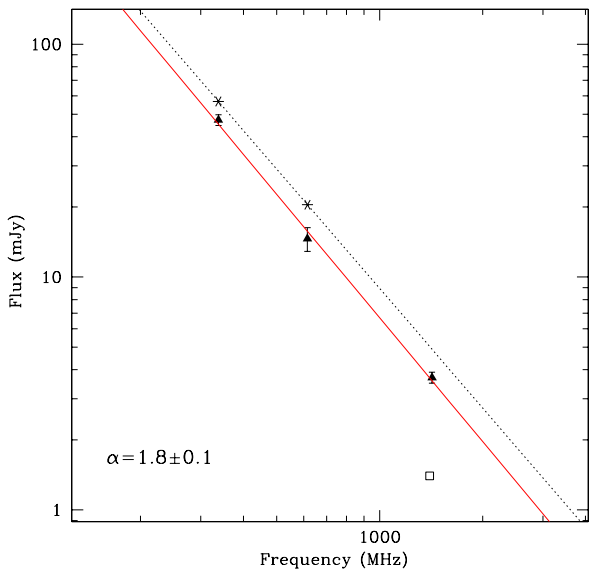}
\caption{Integrated radio spectrum of the halo. The red solid line is the 
linear 
fit to the data (filled triangles), weighted for uncertainties. The dashed 
line connects the two values of flux density including the estimated losses 
at 325 and 610 MHz (stars). The open square is the estimated flux density 
of the radio halo on the NVSS.}
\label{fig:a697_totspec}
\end{figure}

The sources of uncertainty on the radio halo flux densities at 325 MHz and 
1.4 GHz are the calibration errors and the procedure of point source
subtraction, which we estimate sum up to $\sim$ 5\%.
The flux density value at 610 MHz is affected by a higher 
uncertainty. We performed an accurate check on flux density at 325 MHz,
610 MHz and 1.4 GHz for a number of discrete sources in the central field 
(see Fig. \ref{fig:a697_field}), using three images with the same angular 
resolution as the full resolution at 1.4 GHz 
($\sim 16^{\prime\prime} \times 16^{\prime\prime}$), after correction for the 
corresponding primary beam. We found that the 610 MHz measurements are 
systematically underestimated by $\sim 12$\%; 
after further inspection of the data, we concluded that this is due to 
an amplitude sistematic error. Note that 
large residual calibration errors at 610 MHz were reported also for the
cluster A\,3562 in Giacintucci et al. (\cite{giacintucci05}). 
Considering all this, the 610 MHz flux density value of the radio halo in 
A\,697 given in VGD08 was 
corrected to account for this effect (Table 4 reports the new corrected value).
\\
The integrated radio spectrum of the halo is shown in Fig. 
\ref{fig:a697_totspec}. The red solid line is the linear fit to the data
(filled triangles) weighted for the uncertainties. 
A single power--law fit to the data gives a spectral index 
$\alpha$ = 1.8 $\pm$ 0.1\footnote{Giovannini et al.  
(\cite{giovannini09}) estimated a spectral index 
$\alpha_{\rm 325~MHz}^{\rm 1.4~GHz}$=1.2 using our GMRT 325 MHz flux density
value preliminary presented in Venturi et al. (\cite{venturi09}) and the
1.4 GHz value derived from the original NVSS image. However, as shown in 
{\it Sect. 3.1}, the original NVSS value overestimates the flux density at the
centre of A\,697.}

\begin{table}[htbp]
\caption[]{Flux densites of the radio halo in A\,697.}
\begin{center}
\begin{tabular}{cccc}
\hline\noalign{\smallskip}
$\nu$    & S$_{\nu}$  & HPBW      &  Ref.  \\ 
 (MHz)   & (mJy) & $^{\prime\prime} \times ^{\prime\prime}$ &     \\   
\noalign{\smallskip}
\hline\noalign{\smallskip}
325  & 47.3$\pm$2.7 & $46.8 \times 41.4$ & this work; Fig. \ref{fig:a697_halo}  \\
610  & 14.6$\pm$1.7 & $46.4 \times 35.9$ & VGD08; see also this work Fig. \ref{fig:a697_halo}  \\
1400 & 3.7$\pm$0.3  & $35.0 \times 35.0$ & this work; Fig. \ref{fig:halo_1400} $^{*}$  \\
\hline
\end{tabular}
\end{center}
$^{*}$ From VLA--C archival data (Obs. Id. AJ0252) 
\label{tab:fluxes}
\end{table}
%

\section{Uncertainties in the spectral slope}\label{sec:injections}

The combination of very low brightness emission and large angular size
makes the imaging of faint radio halos a challenging process.
The sampling of the u--v coverage at short spacings is essential for a 
reliable imaging of extended and diffuse radio sources, and this is
particularly relevant for faint radio halos, as is the case of A\,697.
The most serious effect of an inadequate u--v coverage at short baselines 
is the loss of a fraction of the radio halo flux density, which may
affect not only the imaging at an individual frequency, but also the 
integrated spectral index of the source. This effect becomes even more severe
when only few data points over a small frequency range are available.
\\
In this Section we will discuss the role of the u--v coverage in the
GMRT observations of A\,697 and the implications on the spectral steepness
of its giant radio halo.
We will show that part of the total flux density of the radio halo 
could not be accounted for by the GMRT observations. This effect is
more severe at 610 MHz, since Fig. \ref{fig:uvcoverage} clearly shows that 
the inner portion of the u--v coverage is much better
sampled in the 325 MHz observations. Given that an underestimate of the
610 MHz flux density would reflect into a steeper spectrum, it is crucial
that we test the reliability of the spectral index derived in the present
work.

To constrain the spectral steepness of the radio halo, we estimated 
the flux density losses expected both in the 325 and 610 MHz GMRT data
following the \textit{\textquotedblleft fake\textquotedblright radio halos} 
procedure, described in detail in VGD08. 
A ``fake radio halo'' is a model of the radio 
halo brightness profile, which consists of a set of 
optically thin concentric spheres with  different radius and flux density. 
Here we briefly describe the main steps of 
this procedure and the results obtained.

\subsection{The 325 MHz data}

In order to evaluate the flux density loss on the halo at 325 MHz, 
families of fake radio halos were injected in the u--v data. We chose
different sets of values for the total flux density and largest angular scale,
and injected the fake halos in a region of the field free of point sources,
and close enough to the cluster centre to avoid attenuation from the
primary beam.
We thus obtained a new dataset, which we will refer to as (u--v)$_{\rm inj}$.
The LAS of the injected fake radio halos (i.e. the diameter of the 
largest sphere of the model) ranges from 240$^{\prime\prime}$ to 
330$^{\prime\prime}$, corresponding to a linear scale in the range 
1$\div$1.4 Mpc at the redshift of A\,697.
The total flux density values injected, $S_{inj}$, range from 50 to 70 mJy.
We then imaged each new dataset (u--v)$_{\rm inj}$ with the same parameters we 
used to produce the 
final low resolution image of the radio halo (see Fig. \ref{fig:a697_halo}), 
and measured the flux density and LAS in each fake radio halo imaged. 
Note that both the fake and the real radio halos are imaged 
simultaneously, allowing a direct comparison between the two sources. 
\\
In general we found that the component with the largest angular size, i.e. 
the one with lowest surface brightness, is partially lost in the imaging 
process. On the contrary, the central highest surface  brightness regions are 
well imaged. These general results are in agreement with what found
in VGD08.
\newline
An injected fake halo with S$_{\rm inj}$=60 mJy and  
LAS=330$^{\prime\prime}$ best reproduces the emission
observed at the cluster centre, i.e. S$\simeq$ 48 mJy and  
LAS$\simeq 300^{\prime\prime}$ (see Sect. \ref{sec:a697_halo}). Based on 
these tests, we concluded that the expected flux density loss 
of the radio halo at 325 MHz is at most $\sim$20 \%. 
Most of this flux density is lost in the largest and lowest surface brightness
component. 
The LAS of the real halo is $\sim 10$ \% smaller than that injected.

\subsection{The 610 MHz data and the ``revised'' spectrum}

Following the same procedure, we estimated the 
flux density loss 
at 610 MHz. A fake halo with LAS=330$^{\prime\prime}$  (same choice as for 
the injection at 325 MHz) and S$_{\rm inj}$=18  mJy best reproduces the radio 
halo observed at 610 MHz. The resulting differential flux density loss 
between 325 MHz and 610 MHz is $\sim20$ \%. 
Based on these results, we estimated that the intrinsic flux density of the 
radio halo in A\,697  is $\sim 57$ mJy at 325 MHz and $\sim 20$ mJy at 
610 MHz (accounting for the 12\% flux density correction derived in Sect. 3.2).

In Fig. \ref{fig:a697_totspec} we report the integrated spectrum of the halo 
between 325 and 610 MHz taking into account the losses in flux density (the 
dashed line connects the two values, represented by the star symbols). The 
resulting spectral index is $\alpha_{\rm rev}\sim 1.7$. 
We conclude that A\,697 hosts a very steep spectrum radio halo.
The flux density we measured at 1.4 GHz can be considered a lower limit, the
upper one being the extrapolation of the ``revised'' spectrum between 325 MHz 
and 610 MHz. This leads to an ``expected'' flux density value 
S$_{\rm 1.4~GHz}\simeq 5$ mJy over the same region considered in this paper. 
Future high sensitivity observations at this frequency may allow us to 
constrain the high frequency end of the radio halo spectrum.

\section{Discussion}\label{sec:discussion}

We have presented observational evidence that A\,697 hosts a very steep 
spectrum giant radio halo. 
\\
This source has observational properties similar to those of A\,521, which
we consider the prototypical example of ultra steep spectrum radio halos.
In particular, 
it is barely detectable at 1.4 GHz with the current instruments, and 
with the observed values logP$_{\rm 1.4~GHz}$(W/Hz) =23.95 and 
L$_{\rm X[0.1-2.4]~keV}=1.06x10^{45}$ erg s$^{-1}$, it lies 
below the well--known logP$_{\rm 1.4~GHz}$--logL$_{\rm X}$ correlation for giant 
radio halos (see Brunetti et al. \cite{brunetti09} for a recent update). 
The source becomes stronger and considerably more extended at lower frequencies.
This is different from what is observed for those radio halos with 
spectral index $\alpha \sim 1.2 - 1.4$, whose overall morphology
and size do not change appreciably moving down to lower frequencies
(see for instance A\,2744 and A\,2219, Orr\'u et al. \cite{orru07}; 
A\,2319, Feretti et al. \cite{feretti01};
MACS J\,0717.5+3745, van Weeren et al. \cite{vanweeren09} and Bonafede et al.
\cite{bonafede09}; 
RXCJ2003.5--2323, Giacintucci et al. \cite{giaci09}).
Based on literature data, Giovannini et al. (\cite{giovannini09}) reported 
on four more halos with
very steep integrated spectrum; however, the different resolutions and u--v 
coverages, and the lack of accurate subtraction of individual embedded radio 
sources in heterogenous datasets, may affect the shape and spectral index of
the integrated spectra.

The observational connection between major cluster mergers and radio halos is a 
fairly well established result, both on the basis of individual studies 
(examples of recent multiband analysis may be found in 
Govoni et al. \cite{govoni04}; van Weeren et al. \cite{vanweeren09}; 
Bonafede et al. \cite{bonafede09}; Giacintucci et al. \cite{giaci09}) 
and on statistical properties of large samples (see for instance 
Buote \cite{buote01} and VGB08).
On the other hand, it has been recently suggested that the majority of 
radio halos should be generated during more common but less energetic mergers, 
for example between a massive cluster and a much smaller subcluster 
(with mass ratio \gtsim~ 5) or between two similar clusters with mass 
M $\lesssim 10^{15}$M$_{\odot}$, which would trigger the formation of radio 
halos observable only up to few hundred MHz (Cassano et al. \cite{cassano06} 
and Cassano \cite{cassano09}).
So far, these elusive radio halos
have been missed from large surveys mainly due to their steep
spectrum, which requires very sensitive observations at 
frequencies $\nu~\le~1$ GHz. The discovery of the ultra steep spectrum radio
halo in A\,521 (Brunetti et al. \cite{a521nature08}) 
provides observational support in favour of this idea. 

The very steep spectrum radio halo found in A\,697 offers a chance to further
test the hypothesis of a connection between ``low frequency radio halos''
and less energetic cluster mergers.

\subsection{The merger in the cluster A\,697}\label{sec:a697_lit}

Optical and X--ray observations are crucial tools to investigate cluster
dynamics.

A detailed optical analysis of A\,697 was carried out in G06, who found 
that the cluster is significantly far from dynamical 
relaxation. In particular, the observed complex dynamical state of A\,697 
might be explained either as an ongoing process of multiple accretion of small 
($\lesssim 0.2$ Mpc) clumps by a very massive cluster, or as the result of a 
past merger event. The scenario may be even more complicated (major merger
followed by multiple accretion), as suggested by the absence of a cool core 
in this cluster (Bauer et al. \cite{bauer05}). G06 concluded that it is
difficult to provide more details on the type of merger in A\,697, since most
likely we are viewing the system at a small angle to the line of sight.
Analysis of archival {\it Chandra} X--ray observations (G06) revealed the
presence of significant substructure in the core region, with three 
clumps of emission within $\sim$200 kpc from the centre, confirming the
cluster complex dynamical state.

\begin{figure}
\centering
\includegraphics[angle=0,width=8.5cm]{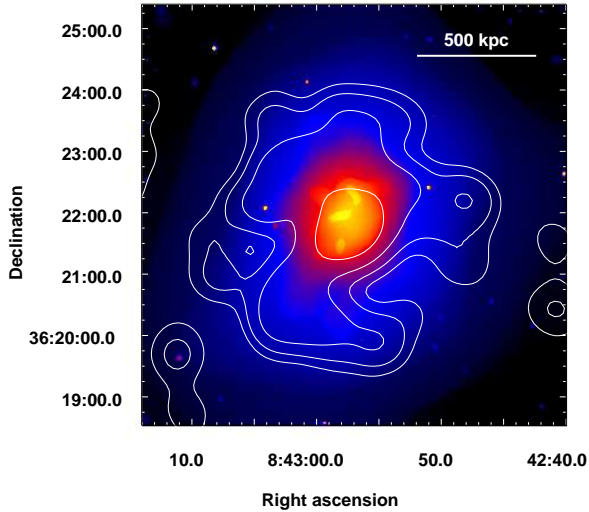}
\caption{Wavelet--reconstructed {\it Chandra} image of A\,697, in the 
0.5--9 keV band. The image is corrected for the exposure and background 
subtracted. Overlaid are the 325 MHz contours of the radio halo. The radio 
image is the same as right panel of Fig. \ref{fig:a697_halo}.}
\label{fig:a697_chandra}
\end{figure}

We re--analysed the public archive {\it Chandra} ACIS--I observation of 
A\,697 (19.8 ks, OBSID 4271, presented also in G06).
In Fig. \ref{fig:a697_chandra} we report a newly wavelet--reconstructed 
{\it Chandra} image of A\,697, with 325 MHz contours of the radio halo 
superposed (same image as right panel of Fig. \ref{fig:a697_halo}).
The X--ray image was obtained applying the wavelet decomposition tool
(Vikhlinin, Forman \& Jones \cite{vikhlinin97}) to the cluster image in the 
0.5--9 keV energy band, divided for the exposure map and background subtracted 
(see Giacintucci \cite{giaciPHD07} and references therein).
\\ 
As clear from Fig. \ref{fig:a697_chandra}, the overall X--ray morphology of the 
cluster is elliptical, and elongated along the  North--North West/South--South 
East (NNW/SSE) direction. 
The extent and shape of the radio emission are in  agreement both 
with the size of the X--ray emitting region and with the slight elongation 
along the NNW/SSE axis. Moreover, the peak of the radio emission covers the 
inner $\sim 200$ kpc of the core region, where the three X--ray subclumps are 
located.
The cluster temperature and hardness ratio distributions were derived in
Giacintucci (\cite{giaciPHD07}), with the calibration files available at the 
time. Gradients between the inner $\sim$200 kpc core region and the 
surrounding areas were found both in the temperature and in the hardness 
ratio, in agreement with the proposed overall picture that A\,697 is 
dynamically unrelaxed. Our analysis favours the multiple merger scenario,
but unfortunately the short exposure time of the {\it Chandra} observations 
does not allow a more quantitative analysis. 

\label{fig:a697_wavopt}

\subsection{Origin of the radio halo}\label{sec:halo_origin}

Theoretically, cosmic ray protons are expected to be the dominant
non--thermal particle component in the ICM, due to their long life--time
(see Blasi et al. \cite{blasi07} for a recent review). However, current 
observations in
different bands constrain the energy density of the non--thermal component
to $<$ 10\% of the thermal gas in the central Mpc region of galaxy 
clusters (e.g. Brunetti et al. \cite{brunetti07}; Churazov et al. 
\cite{churazov08}; Aharonian et al. \cite{aharonian09a} and \cite{aharonian09b};
Lagan\'a et al. \cite{lagana10}).

\noindent
As we mentioned in the introduction,
it is presently believed that two main physical processes may  
contribute to the origin of the extended diffuse
synchrotron emission in giant radio halos: the injection of
secondary electrons through proton--proton collisions
(hadronic models, e.g. Dennison 1980; Blasi \& Colafrancesco \cite{blasi99};
Pfrommer \& Ensslin \cite{pfrommer04}), and the in--situ re--acceleration of
relativistic electrons by MHD turbulence generated in the
ICM during cluster-cluster mergers (re--acceleration models, 
Brunetti et al.~2001, 2004; Petrosian 2001; Fujita et al. \cite{fujita03};
Cassano \& Brunetti \cite{cassano05}).

\noindent
Radio halos with very steep spectrum, i.e. $\alpha \geq 1.6$, are suitable 
targets to constrain these models and favour a turbulent 
re--acceleration scenario (Brunetti et al. \cite{a521nature08}): 
in fact, in the context of the hadronic scenario, clusters 
hosting these radio halos must contain a very large population of cosmic 
ray protons.
The observations presented in this paper suggest that Abell 697 hosts
a very steep spectrum giant halo and this allows a prompt test of the
hadronic model.

In this Section we assume that the giant radio halo in A\,697 is of hadronic 
origin and discuss the consequences of this hypothesis on the physical 
properties of the ICM.

\noindent
The decay chain that we consider for the injection
of secondary particles in the ICM due to p-p
collisions is (Blasi \& Colafrancesco \cite{blasi99}):

$$p+p \to \pi^0 + \pi^+ + \pi^- + {\ldots} $$
$$\pi^0 \to \gamma \gamma$$
$$\pi^\pm \to \mu + \nu_\mu ~~~ \mu^\pm\to e^\pm \nu_\mu \nu_e.$$

\noindent
that is a threshold reaction that requires protons with kinetic
energy larger than $T_p \approx 300$ MeV. 

Under the assumption that secondary electrons are not accelerated by other 
mechanisms, their spectrum approaches a stationary distribution due to
the competition between injection and energy losses 
(Dolag \& Ensslin \cite{dolag00}):

\begin{equation}
N_e^{\pm}(p)=
{1 \over
{\Big| \left( {{dp}\over{dt}} \right)_{\rm loss} \Big| }}
\int_{p}^{p_{\rm max}}
Q_e^{\pm}(p) dp.
\label{sec_stat}
\end{equation}

\noindent
where $Q_e^{\pm}$ is the injection rate of secondary electrons
and 
radiative losses, that dominate for $\gamma > 10^3$ electrons in the ICM, 
are (Sarazin \cite{sarazin99}) :

\begin{equation}
\Big| \left( {{ d p }\over{d t}}\right)_{\rm loss} \Big|
\simeq 3.3 \times 10^{-32} 
\big( {{p/m_e c}\over{300}} \big)^2 \left[ \left( {{ B_{\mu G} }\over{
3.2}} \right)^2 + (1+z)^4 \right]. 
\label{loss}
\end{equation}

\noindent
We follow standard formulae to calculate the injection rate of secondary
electrons (Moskalenko \& Strong \cite{moskalenko98}; 
Blasi \& Colafrancesco \cite{blasi99}; Brunetti \& Blasi \cite{brunetti05}) 
and use Dermer's
fitting formulae for the inclusive p-p cross section (Dermer \cite{dermer86}) 
which allow to describe separately the rates of generation of $\pi^{-}$, 
$\pi^{+}$ and $\pi^{o}$.

\noindent
We assume a power law distribution of relativistic
protons, $N_p(p) = K_p p^{-s}$, in which case 
the spectrum of secondaries at high energies, 
$\gamma > 10^3$, is $N_e(p) \propto p^{-(s+1)} {\cal F}(p)$,
where ${\cal F}$ accounts for the Log--scaling 
of the p-p cross section at high energies and makes the spectral shape 
slightly flatter than $p^{-(s+1)}$ (Brunetti \& Blasi \cite{brunetti05}; 
Brunetti \cite{gb09}).

\noindent
The synchrotron emissivity from secondary e$^{\pm}$ is also obtained via 
standard formulae (Rybicki \& Lightman \cite{rybichi79}; see also
Dolag \& Ensslin \cite{dolag00}) :

\begin{eqnarray}
J_{syn}(\nu) = \sqrt{3} {{e^3}\over{m_e c^2}} B 
\int_0^{\pi/2} d\theta sin^2\theta \int dp N_e(p) 
F\big( {{\nu}\over{\nu_c}} \big)
\nonumber\\
\propto n_{th} \epsilon_{CR}  
{{B^{1+\alpha} }\over{B^2 + B_{cmb}^2}} \nu^{-\alpha}
\label{jsyn}
\end{eqnarray}

\noindent 
where

\begin{equation}
\epsilon_{CR} = K_p \int p^{-s} T_p dp
\end{equation}

\noindent
is the energy density of cosmic ray protons ($T_p$ is the
kinetic energy of protons),  
$F$ is the synchrotron Kernel (Rybicki \& Lightman \cite{rybichi79}), 
$\nu_c$ is the critical frequency, and 
$\alpha \simeq s/2 -\Delta$, $\Delta \sim 0.1-0.15$ due to the logarithmic
scaling of the cross section (eg., Brunetti 2009).
We adopt a value of the spectral index of the halo $\alpha = 1.75$
which is consistent with our observational findings and 
implies a spectral slope of the cosmic ray protons $s=3.8$.

The diffuse emission of the radio halo is centered on the cluster 
core region and grossly resembles the X-ray morphology both in terms of 
extension and elongation (Fig.~\ref{fig:a697_chandra}).
\noindent
The radio profile at 325 MHz, however is flatter than the X-ray 
profile,
it drops by only a factor 5--6 at distance of 2.5--3 core radii 
($\sim$ 500 kpc), suggesting a broad spatial distribution of the 
relativistic particles.
Thus, we use a $\beta$--model for the spatial distribution of the 
cluster thermal gas (with parameters taken from Bonamente et al. 
\cite{bonamente06}:
$\beta$=0.6, 
$n_o=9.8\times10^{-3}$cm$^{-3}$, 
$r_c=42^{\prime\prime}$, i.e. $\approx$ 180 kpc, and 
gas temperature kT = 10 10keV)
and adopt a model 
({\it Model} 1) of the distribution of relativistic components in
Abell 697 where the energy density of both cosmic ray protons and
magnetic field are {\it constant} with radius up to a distance $= 2.5 \, r_c$,
and scale with that of the thermal gas at larger distances.
This model gives a very good representation of the observed halo-radial profile
up to a distance of about 2.5$~r_c$ and, within the uncertainties, it is also 
in line with the observed profile at large distance (regardless of the 
strength of the magnetic field at $\leq 2.5 \, r_c$ distance).

\begin{figure}
\centering
\includegraphics[angle=0,width=8.5cm]{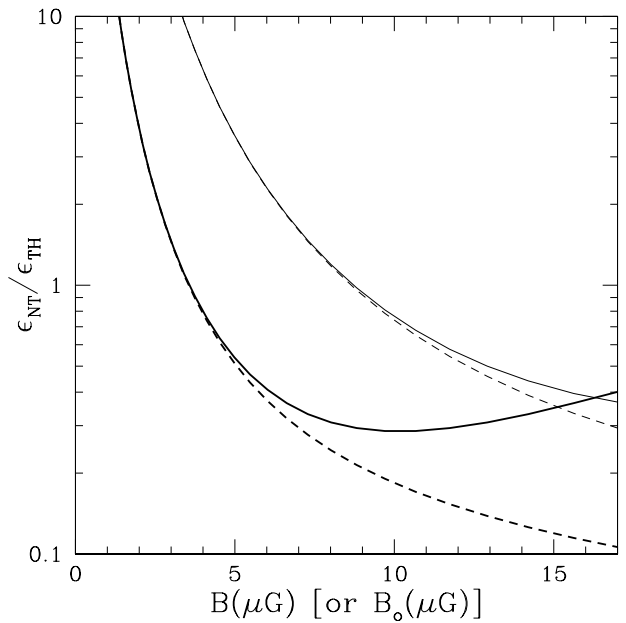}
\caption{The ratios between the energy densities
of relativistic protons and thermal ICM (dashed lines) and
between non-thermal (relativistic protons and magnetic field) and
thermal ICM (solid lines) are shown as a function of magnetic field.
All energy densities are calculated within a distance $r \sim 3 \, r_c$,
i.e. roughly the region filled by the radio halo.
Thick lines refer to {\it Model} 1, in which case $B$ is that
within $2.5 \, r_c$.
Thin lines are obtained by assuming that the magnetic field 
energy-density scales with the thermal one, $B= B_o (n_{th} / n_o)^{1/2}$
(e.g. Govoni \& Feretti \cite{fg04}; with thermal quantities taken
from Bonamente et al. \cite{bonamente06}),
and by leaving the radial profile of 
relativistic protons free to vary with distance 
(up to $r \sim 2.5 r_c$ while at larger distances it is assumed to scale 
with thermal one as in {\it Model} 1) in order to match the 
observed synchrotron profile; in this latter case the x--axis shows the
central field value, $B_o$.}
\label{fig:model}
\end{figure}

\noindent
In Fig. \ref{fig:model} we show the ratio of the energy density 
of non--thermal (relativistic protons and magnetic field) and thermal 
components in Abell 697 that is required assuming a hadronic origin of 
the radio halo.
We find that for $B \leq 5 \mu$G relativistic protons are required 
to store an energy comparable to (or larger than) that of the thermal ICM.
The non--thermal energy content reaches a minimum for $B \approx 10 \mu$G, 
$\epsilon_{NT} \approx 1/3 \epsilon_{TH}$, implying 
an important dynamical contribution of the non--thermal components
in the cluster. 
In addition, we point out that {\it Model} 1 provides only a lower limit 
to the energy of the non--thermal components for two main reasons :

\begin{itemize}

\item{1)} as soon as we 
include also the 
contribution from the tail of the
proton energy distribution at sub--relativistic energies (E$<$ 1 GeV), 
the required energy budget is much larger than that in 
Fig.~\ref{fig:model} due to the steep proton spectrum, 
$\epsilon_{CR} \propto p_{min}^{-s+3}$.

\item{2)} the rather unphysical assumption in {\it Model} 1 that 
the energy density of the 
non-thermal components is {\it constant} with radius (up $2.5 \, r_c$) 
yields a lower limit to the energy density of the cosmic ray protons that
is required by the hadronic scenario.
Indeed as soon as the magnetic field strength in Abell 697 is allowed to 
decrease with distance from the cluster centre, the energy density of 
relativistic protons must increase with radius and their total energy budget 
increases (Fig.~\ref{fig:model}).

\end{itemize}

We conclude that the very large energetics necessary for the non--thermal 
components assuming a hadronic origin of the halo disfavours this scenario.

\section{Summary and conclusions}\label{sec:results}

In this paper we presented 325 GMRT high sensitivity observations of
the giant radio halo in A\,697, and performed an accurate study of
its integrated spectrum.

The largest extent of the radio halo at 325 MHz is $\sim 5^{\prime}$ 
(corresponding to 1.3 Mpc h$_{\rm 70}^{-1}$), and it is considerably 
larger than at higher frequencies. 
The radio spectrum, determined with three data points in
the frequency range 325 MHz--1.4 GHz, has a spectral index $\alpha_{\rm
325~MHz}^{\rm 1.4~GHz} = 1.8\pm 0.1$.
An accurate analysis of the fraction of flux density that might be 
missed in our observations due to the u--v coverage at short spacings 
led to an estimated spectral index  $\alpha_{\rm rev}$=1.7. 
We thus conclude that A\,697 hosts a very steep spectrum radio halo.

A qualitative analysis of the X--ray emission from the ICM suggests that
this cluster is unrelaxed. Comparison with the 325 MHz emission from
the radio halo shows that the non--thermal and thermal emissions have 
a similar morphology.
Though at a qualitative level, the observations are in agreement with
the idea that A\,697 is the result of a multiple merger.

Similarly to the case of A\,521, we showed that the very 
steep spectrum of the halo disfavours a hadronic origin, which 
would require an unplausibly large energy budget.
On the other hand, models based on turbulent acceleration for the origin of 
radio halos predict a large number of halos in the Universe with very steep 
spectrum. This can be tested as soon as LOFAR and LWA will become operational.

\begin{acknowledgements}
We thank the staff of the GMRT for their help during the observations. GMRT
is run by the National Centre for Radio Astrophysics of the Tata Institute of
Fundamental Research. 
This research is partially funded by INAF and ASI through grants 
PRIN--INAF 2007, PRIN--INAF 2008 and ASI--INAF I/088/06/0. 
This research has made use of the NASA/IPAC Extragalactic Database 
(NED) which is operated by the Jet Propulsion Laboratory, 
California Institute of Technology, under contract with the 
National Aeronautics and Space Administration.

 \end{acknowledgements}

\end{document}